\documentclass[fleqn,twoside]{article}
\usepackage{espcrc2}
\usepackage{url}
\usepackage{graphicx}
\def\lesssim{\mathrel{\mathpalette\vereq<}}
\def\gtrsim{\mathrel{\mathpalette\vereq>}}
\makeatletter
\def\vereq#1#2{\lower3pt\vbox{\baselineskip1.5pt \lineskip1.5pt
\ialign{$\m@th#1\hfill##\hfil$\crcr#2\crcr\sim\crcr}}}
\makeatother

\title{Alternatives to Seesaw}

\author{Hitoshi Murayama\address[IAS]{School of Natural Sciences,
    Institute for Advanced Study\\ 
    Princeton, NJ 08540, USA}%
  \thanks{On leave of absence from \it Department of Physics,
    University of California, Berkeley, CA 94720.}}
       
\begin{document}

\begin{abstract}
  The seesaw mechanism is attractive not only because it ``explains''
  small neutrino mass, but also because of its packaging with the
  SUSY-GUT, leptogenesis, Dark Matter, and electroweak symmetry
  breaking.  However, this package has the flavor, CP, and gravitino
  problems.  I discuss two alternatives to the seesaw mechanism.  In
  one of them, the anomaly-mediated supersymmetry breaking solves
  these problems, while predicts naturally light Dirac neutrinos.  In
  the other, the light Majorana neutrinos arise from supersymmetry
  breaking with right-handed neutrinos below TeV, and the Dark Matter
  and collider phenomenology are significantly different.
  \vspace{1pc}
\end{abstract}

\maketitle

\section{Introduction}

We are here to celebrate 25 years of the seesaw mechanism
\cite{seesaw}.  This attractive mechanism is meant to explain why the
neutrino masses are finite, yet tiny compared to all the other
elementary fermion masses.  Tsutomu Yanagida and Pierre Ramond in the
audience pioneered this mechanism from the point of view of the
$SO(10)$ grand unified theories.  It predicts the mass of the
right-handed neutrinos $N$ at the GUT-scale $M_R \sim M_{GUT} \gg 2
\times 10^{16}$~GeV$\gg v = 176$~GeV, and the light neutrino mass is
suppressed tremendously as $m_\nu \sim v^2 / M_{GUT} \simeq 1$~meV.

Actually, what most of us find attractive is not just the mechanism to
suppress the neutrino mass, but rather the whole package of the grand
unification \cite{Georgi:1974sy}, the supersymmetry to stabilize the
hierarchy
\cite{Veltman:1980mj,Dimopoulos:1981au,Witten:1981nf,Dine:1981za,Dimopoulos:1981yj,Dimopoulos:1981zb,Sakai:1981gr},
the seesaw mechanism to suppress the neutrino mass, and the (thermal)
leptogenesis that explains the baryon asymmetry of the universe and
hence ``why we exist'' in terms of the out-of-equilibrium decay of the
right-handed neutrinos \cite{Fukugita:1986hr}.  Furthermore, the whole
package comes with all the usual goodies of the supersymmetry, namely
the Lightest Supersymmetric Particle (LSP) as a natural candidate for
the cosmological Dark Matter \cite{Goldberg:1983nd,Ellis:1983ew} and
radiative breaking of the electroweak symmetry
\cite{Inoue:1982ej,Inoue:1982pi,Alvarez-Gaume:1983gj,Ibanez:1983di},
potentially connected to the superstring theory.

\begin{figure}[htbp]
  \centering
  \includegraphics[width=0.6\columnwidth]{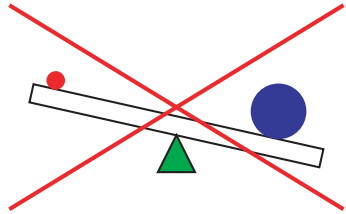}
  \caption{Talking about alternatives to the seesaw mechanism.}
  \label{fig:noseesaw}
\end{figure}

Why talk about alternatives then?  Even though this package is
compelling and attractive, it has problems.  First of all, the
supersymmetry comes with the flavor and CP problems due to the new
particles and their flavor- and CP-violating parameters below the TeV
scale (see, {\it e.g.}\/, \cite{Murayama:2000fm}).  In view of the
string unification, the GUT-scale, which is supposed to explain the
smallness of the neutrino mass, is put in by hand arbitrarily and is
not explained.  Furthermore, the proton decay had not been seen at the
predicted rate and the naive SUSY-GUT appears to be in trouble
\cite{Murayama:2001ur}.  Concerning the cosmology, the thermal
leptogenesis requires the (lightest) right-handed neutrino mass to be
$M_R \gtrsim 10^{9}$~GeV, and hence the reheating temperature after
the inflation $T_{RH} \gtrsim 10^{10}$~GeV
\cite{Buchmuller:2004nz}.\footnote{Several ways out of this limitation
  had been discussed in the literature, such as resonant leptogenesis
  \cite{Pilaftsis:2003gt} or soft leptogenesis \cite{Grossman:2003jv}.
  Another possibility is the coherent oscillation of right-handed
  sneutrinos \cite{Murayama:1993em,Hamaguchi:2001gw} which may even be
  the inflaton itself \cite{Murayama:1992ua,Ellis:2003sq}. } On the
other hand, the gravitinos, expected at $m_{3/2} \simeq
100$---1000~GeV, are produced thermally at high temperatures and decay
late, jeopardizing the success of the Big-Bang Nucleosynthesis (BBN)
theory.  In fact, the hadronic decays of the gravitinos cause so much
trouble that the reheating temperature must be kept below $T_{RH}
\lesssim 10^6$~GeV to suppress the abundance of the gravitinos
sufficiently \cite{Kawasaki:2004yh,Kawasaki:2004qu}.  Therefore the
leptogenesis and supersymmetry appear to be in conflict.  Finally, the
seesaw mechanism itself is very difficult to test.  In particular, the
thermal leptogenesis prefers neutrino mass eigenvalues
$m_{\nu_{1,2,3}} \lesssim 0.1$~eV \cite{Buchmuller:2004nz}, and hence
the detection of the neutrinoless double beta decay is far from
trivial.

Therefore I believe it is worthwhile discussing alternatives to the
by-now-standard seesaw mechanism.  I will discuss two such
possibilities. 

In the first one, the neutrino mass is suppressed by the hierarchy
between the Planck-scale and the supersymmetry breaking scale.
Normally this hierarchy would give a too small neutrino mass; however
in the anomaly-mediated supersymmetry breaking
\cite{Randall:1998uk,Giudice:1998xp}, the superparticle masses are
suppressed by loop-factors, and correspondingly the source of the
supersymmetry breaking, {\it i.e.}\/, the gravitino mass, is enhanced
relative to the TeV-scale.  It predicts the neutrino mass of $m_\nu
\sim {\rm 100 TeV}/M_{Pl} \sim 10$~meV, much closer to the data of
8--50~meV than the conventional prediction from the SUSY-GUT $m_\nu
\sim 1$~meV \cite{Arkani-Hamed:2000xj}.  In addition, the
anomaly-mediated supersymmetry breaking automatically solves the SUSY
flavor problem.  In this implementation, the neutrinos are Dirac
fermions.  If we are lucky, the long-baseline neutrino oscillation
experiments will tell us that the neutrino spectrum is inverted, while
the negative search for the neutrinoless double beta decay will set
the limit $|\langle m_\nu\rangle_{ee}| < 0.01$~eV.  If this happens,
we will establish the Dirac nature of neutrinos, giving a clear
preference to a scenario of this type over the conventional seesaw.
Despite the conserved lepton number, it is possible to have ``Dirac
leptogenesis'' explaining the cosmic baryon asymmetry
\cite{Dick:1999je,Murayama:2002je}.

%% In the second possibility, the neutrino mass arises again from the
%% supersymmetry breaking effect.  

\section{Consistent Anomaly Mediation}

In this section, we present a framework where SUSY flavor and CP
problems are solved, as well as the cosmological gravitino problem.
The neutrinos are naturally light but Dirac, yet the leptogenesis is
possible.  It relies heavily on the anomaly-mediated supersymmetry
breaking.  

\subsection{Flavor and CP Problems}

\begin{figure}[htbp]
  \centering{
    \includegraphics[width=\columnwidth]{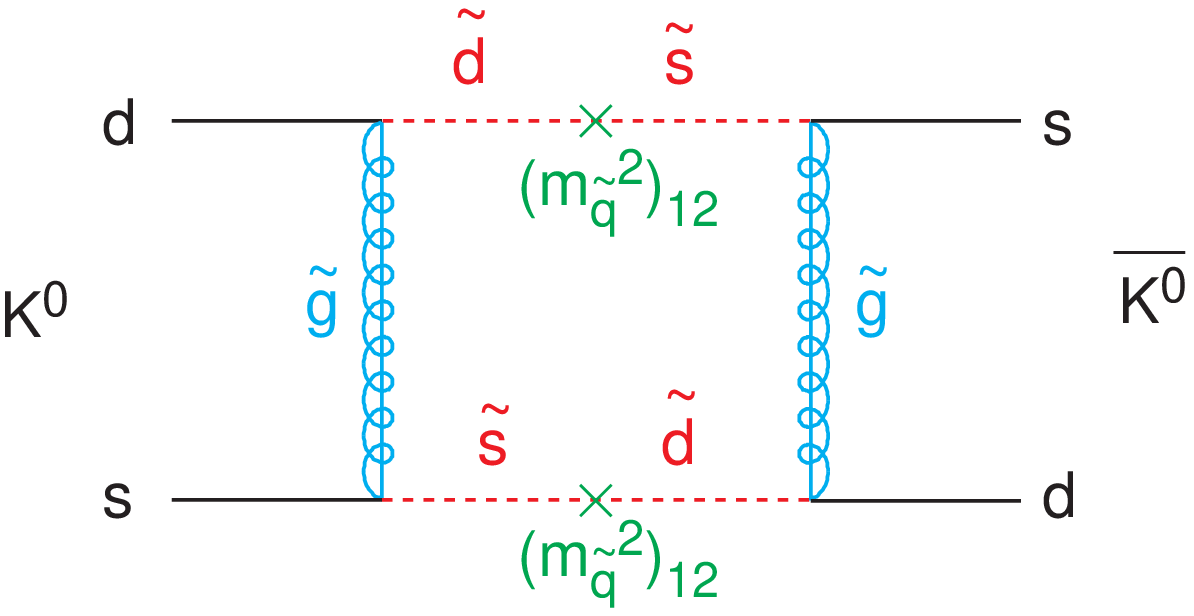}
    \includegraphics[width=\columnwidth]{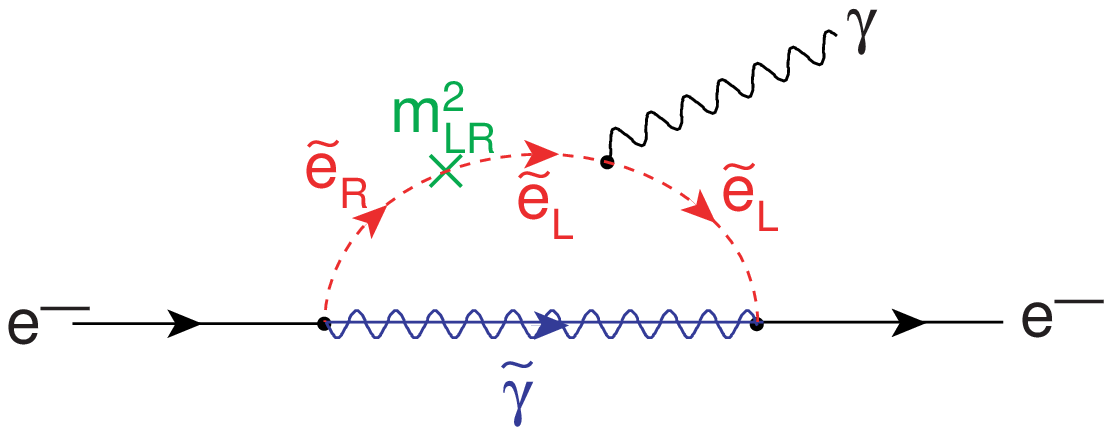}
  }
  \caption{The potentially too large SUSY contribution to the neutral
  kaon mixing and electron electric dipole moment.}
  \label{fig:problems}
\end{figure}

It is well-known that generic supersymmetry breaking effects would
induce unacceptably large flavor-changing effects as well as
CP-violating effects.  For example, the diagrams in
Fig.~\ref{fig:problems} can induce too-large contribution to neutral
kaon mixing or electron electric dipole moment (EDM).  The vertices
indicated by green crosses must be extremely suppressed in order to
satisfy the experimental constraints, down to $10^{-4}$ of the natural
size in some cases.

Many proposals exist that avoid these serious problems in
supersymmetry.  The popular frameworks are gauge mediation
\cite{Dine:1993yw,Dine:1994vc,Dine:1995ag} or gaugino mediation
\cite{Kaplan:1999ac,Chacko:1999mi} of supersymmetry breaking.  In
these frameworks, the masses and mixings of superpartners are induced
by the standard-model gauge interactions, and hence are flavor neutral
and do not induce flavor-changing effects.  With some extra work, they
can be made also real, hence CP-preserving.  In both cases, the idea
is to {\it overcome}\/ the ``bad'' supergravity induced supersymmetry
breaking effects by the ``good'' gauge induced effects.  Namely
actively trying to induce large effects is the approach.

However, the gauge mediation predicts a light gravitino, between
100~keV and 10~GeV, which has also a severe cosmological limits from
the overclosure and the BBN
\cite{Moroi:1993mb,deGouvea:1997tn}.\footnote{Note, however, some
  small portions of the parameter space allow viable yet interesting
  cosmology of gravitino dark matter, studied recently for example in
  \cite{Fujii:2002fv,Feng:2003uy,Buchmuller:2004tm,Feng:2004zu}.}  The
gaugino mediation requires a rather high energy scale for the
mediation, typically above the GUT-scale, so that the slepton masses
are enhanced enough by the renormalization-group running.  It
therefore leaves the concern that some flavor physics below the
mediation scale, such as right-handed neutrinos in the seesaw
mechanism, induce flavor-dependent effects despite the initial
flavor-blind boundary conditions.  Moreover, neither framework seems
to provide obvious connections to neutrino physics.  I will not
discuss them any further here.

The alternative approach is the anomaly mediation.  In this framework,
one actively tries {\it not}\/ to induce supersymmetry breaking
effects, by making the sector responsible for supersymmetry breaking
as ``sequestered'' from the standard model as possible.  I call this
approach ``Zen of supersymmetry breaking:'' You try not to mediate,
and you mediate good ones.  One such sequestering mechanism is the
physical separation of two sectors on separate points in the extra
dimensions \cite{Randall:1998uk}, another is to use ``conformal
sequestering'' motivated by the AdS/CFT correspondence
\cite{Luty:2001jh,Luty:2001zv}, where the unwanted direct coupling of
the two sectors is suppressed by a near conformal dynamics of the
supersymmetry breaking sector.

\begin{figure}[htbp]
  \centering
  \includegraphics[width=0.6\columnwidth]{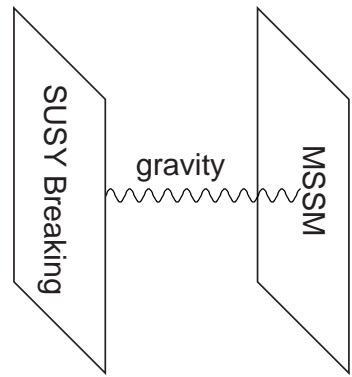}
  \caption{Physical separation of the standard model and the sector
  that breaks supersymmetry in extra dimensions that lead to anomaly
  mediation as the dominant contribution to the supersymmetry
  breaking.} 
  \label{fig:AMSB}
\end{figure}

The point is simply that the gravity always exists.  And the gravity
couples to energy.  All parameters in the theory that have dimensions
of energy receive supersymmetry breaking effects through gravity,
while dimensionless parameters do not.  All parameters in the standard
model except for one, namely the Higgs negative mass squared, are
dimensionless, and they do not receive supersymmetry breaking effects.
However, the dimensionless parameters are actually secretly
dimensionful due to the running of the coupling constants.  They do
depend on mass scales.  Because of the dependence on the mass scales,
namely the conformal anomaly, the supersymmetry breaking effects
appear proportional to the beta functions $\beta_A(g_A)=\mu
\frac{d}{d\mu} g_A$ of the running gauge coupling constants or the
anomalous dimension factors $\gamma_i = - \frac{1}{2} \mu
\frac{d}{d\mu} \log Z_i$ which make the Yukawa couplings run.  The
result is
\begin{eqnarray}
  A_{ijk} &=& - \lambda_{ijk} ( \gamma_i + \gamma_j + \gamma_k ) m_{3/2}\\
  \tilde{m}_i^2 &=& \frac{1}{2} \mu\frac{d \gamma_i}{d \mu} m_{3/2}^2\\
  M_A &=& \frac{\beta_A}{g_A}m_{3/2}\ , \label{eq:soft}
\end{eqnarray}
a consequence of the {\it superconformal anomaly}\/, hence the name.

The most remarkable point about this result is that it is completely
``UV-insensitive.''  The predictions are given in terms of the
interactions at the energy scale of interest only; whatever happens at
much higher energy scales does not affect the result.  This is in
stark contrast to other scenarios, where any physics below the scale
of mediation would correct and modify the supersymmetry breaking
effects at the electroweak scale through renormalization, and hence
UV-sensitive.  In the supersymmetric Standard Model, the only
flavor-dependence appears through the Yukawa couplings, which are
practically negligible for the first- and second-generation particles.
Therefore the supersymmetry breaking effects are flavor-blind, except
for the third-generation particles.  It automatically solves the
flavor problem in supersymmetry.

I cannot overemphasize how incredible the UV-insensitivity is.  For
example, one can verify this fact by writing down a model with extra
heavy particles and carefully integrate them out later.  The
supersymmetry breaking effects above the mass thresholds of heavy
particles are complicated and flavor-dependent, but once they are
integrated out, the threshold corrections modify the low-energy
supersymmetry breaking effects in such as way that they become
flavor-blind \cite{Giudice:1998xp,Boyda:2001nh}.  This is the opposite
from the other ``flavor-blind'' mediation mechanisms.  They set the
boundary conditions in the UV to be flavor-blind, and the flavor
physics below that scale screw up the original flavor blindness.  In
anomaly mediation, the supersymmetry breaking effects at high energies
are complicated and flavor-dependent; yet after integrating out heavy
degrees of freedom, the flavor-dependence magically gets canceled out.

\subsection{Gravitino Problem}

\begin{figure}[htbp]
  \centering{
    \includegraphics[width=0.95\columnwidth]{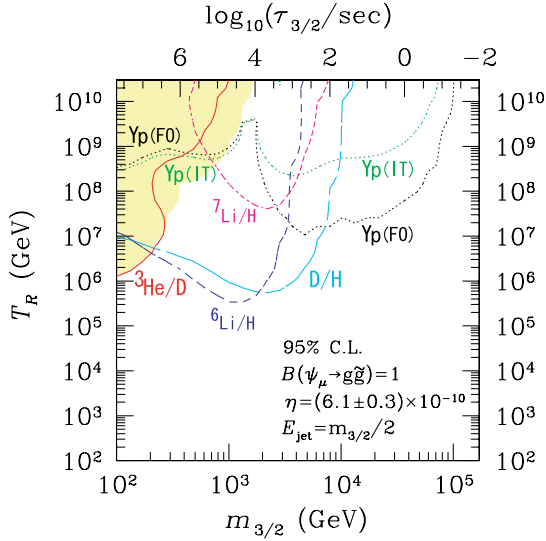}
  }
  \caption{The constraint on the reheating temperature from the
  compatibility of the Big-Bang Nucleosynthesis and the hadronic decay
  of gravitinos produced after the inflation \cite{Kawasaki:2004qu}.} 
  \label{fig:gravitino}
\end{figure}

The immediate gratification of anomaly mediation is the resolution of
the cosmological gravitino problem.  Once supersymmetry is assumed,
there is a superpartner of the gravitino, the gravitino $\tilde{G}$.
Its interaction is only gravitational and hence weak.  Therefore its
lifetime is expected to be long, $\tau_{3/2} \propto
M_{Pl}^2/m_{3/2}^3$, and typically decays after the BBN and dissociate
some of the light elements already synthesized by that time.  This
would destroy the agreement between the BBN theory and the observed
light element abundances.  If the gravitinos are as populous as other
particle species, it is a disaster.  We have to assume that the
gravitinos had been wiped out by the cosmological inflation.  However
the hot gas in the early universe produces gravitinos from the
scattering process such as $g g \rightarrow \tilde{g} \tilde{G}$.  Its
abundance is given approximately by $n_{3/2}/s \simeq 10^{-12}
T_{RH}/10^{10}~\mbox{GeV}$, larger for the higher reheating
temperature after the inflation.  Therefore the success of the BBN
theory places an upper limit on $T_{RH}$.  In particular, when the
gravitinos decay dominantly into hadrons, the constraints were found
to be particularly strong \cite{Kawasaki:2004yh}.  The limit is shown
in Fig.~\ref{fig:gravitino} as a function of the gravitino mass.  For
a TeV gravitino, the upper limit is $T_{RH} \leq 300$~TeV, making the
thermal leptogenesis very difficult.

In anomaly mediation, this problem is basically solved automatically.
Because the supersymmetry breaking effects are induced from the
superconformal anomaly, and hence loop suppressed, the gravitino mass
is enhanced relative to the superpartner masses, $m_{3/2} \simeq
(4\pi)^2 m_{\tilde{q}} \simeq 100$~TeV.  Such a heavy gravitino makes
it decay well before the BBN, making it safe.  There is an additional
constraint that the LSP in the decay product of the gravitinos should
not overclose the universe \cite{Kawasaki:1994af},
\begin{equation}
  \label{eq:TRH}
  T_{RH} \leq 3\times 10^{10}~{\rm GeV} \left( \frac{m_{LSP}}{\rm
  100~GeV}\right).
\end{equation}
Note that this is a much weaker constraint than those in
Fig.~\ref{fig:gravitino}.

\subsection{Slepton Mass Problem}

However, there is a serious problem with the anomaly mediation.  It is
too predictive!  It has only one free parameter, the gravitino mass
$m_{3/2}$, in predicting all superpartner masses.  In particular, the
slepton mass-squared is predicted to be {\it negative}\/, breaking the
electromagnetism and making the photon massive.  You wouldn't be able
to see my slides!\footnote{Pierre Ramond pointed out that it may be a
  good thing.  {\tt \^{ }\_\^{ };}} Phenomenologically, the framework
is DOA: ``dead on arrival.''  

There had been many fixes proposed to this problem
\cite{Pomarol:1999ie,Chacko:1999am,Katz:1999uw,Bagger:1999rd,Kaplan:2000jz,Chacko:2001jt,Okada:2002mv,Nelson:2002sa,Anoka:2003kn}.
One of the simplest is to add a universal scalar mass.  More elaborate
one uses a ``non-supersymmetric threshold,'' namely a flat direction
that acquires a large expectation value due to supersymmetry breaking
effects.  The unfortunate aspect of all these proposals is that they
fix the problem by abandoning the UV insensitivity one or the other
way.  They are therefore not immune from the flavor problem anymore.
One possibility to make the slepton masses-squared positive without
abandoning the UV insensitivity is to introduce new Yukawa
interactions for leptons; within the supersymmetric Standard Model,
only such possibility is the $R$-parity violation
\cite{Allanach:2000gu}, but it breaks the lepton number too much (and
hence too large neutrino mass).  Moreover the UV insensitivity does
not guarantee the absence of the flavor problem either because the
$R$-parity violating couplings are flavor-dependent.  Specific choices
need to be made to avoid this problem.

There is, fortunately, a simple way to make the slepton masses
positive while maintaining the UV insensitivity
\cite{Arkani-Hamed:2000xj}, by adding $D$-term contributions to the
scalar masses $m_i^2 \rightarrow m_i^2 + q_i D$ \cite{Jack:2000cd}.
The UV insensitivity is preserved when the U(1) symmetry is anomaly
free with respect to the standard model gauge group
\cite{Arkani-Hamed:2000xj}.  In the MSSM there are two candidates of
the anomaly free U(1) symmetries, i.e., U(1)$_Y$ and U(1)$_{B-L}$, and
those $D$-term contributions are sufficient to resolve the tachyonic
slepton problem.  Although U(1)$_Y$ is unbroken above the electroweak
scale, the kinetic mixing between U(1)$_{B-L}$ and U(1)$_Y$ induces a
$D$-term for U(1)$_Y$, once a $D$-term for U(1)$_{B-L}$ is generated.

\subsection{Neutrino Mass}

Now we turn our attention to the neutrino mass in this framework.  In
order to add the $U(1)_{B-L}$ $D$-term and maintain the UV
insensitivity, $U(1)_{B-L}$ remains as a global symmetry of the
theory.  It then forbids the Majorana mass of neutrinos, and hence the
standard seesaw mechanism cannot be used.\footnote{There is, however,
  a possible compromise.  The standard seesaw mechanism breaks
  $U(1)_{B-L}$ explicitly and hence induces UV sensitivity, but only
  at the one-loop level without the usual logarithmic enhancement
  factors.  This is an attractive possibility that allows for the
  standard thermal leptogenesis while keeping the lepton-flavor
  violation at minimum \cite{Ibe:2004tg}.}

It turns out that this is not a problem but rather a virtue.  Having
imposed $U(1)_{B-L}$, the right-handed neutrinos $N$ must be light
without their usual Majorana mass.  Clearly $O(1)$ neutrino Yukawa
couplings must be forbidden to avoid too heavy neutrinos.  It can be
done, for example, by a $U(1)_R$ symmetry with charge $+2/3$ for $Q$,
$L$, charge $+1/3$ for $U$, $D$, and $E$, charge $+1$ for $H_u$ and
$H_d$, and charge $-5/3$ for $N$.  On the other hand, the K\"ahler
potential term
\begin{equation}
  \label{eq:Kahler}
  \int d^4 \theta \frac{1}{M_{Pl}} L H_u N
\end{equation}
is allowed.  Normally, such a term is discarded because it is a total
derivative within the global supersymmetric theory.  However, the
supersymmetry breaking effects appear for any dimensionful couplings
in the theory, and this term is dimensionful due to the Planck-scale
suppression.  It can be represented by the Weyl compensator $\Phi = 1
+ \theta^2 m_{3/2}$, which appears in the above term as
\begin{eqnarray}
  \label{eq:Kahler2}
  \lefteqn{
    \int d^4 \theta \Phi^* \Phi \frac{1}{M_{Pl}} L H_u N } \nonumber \\
  &&= \int d^2 \theta \frac{m_{3/2}}{M_{Pl}} L H_u N 
  + O(m_{3/2}^2)
\end{eqnarray}
The first term is extremely interesting: the neutrino Yukawa coupling
is generated but is suppressed by $m_{3/2}/M_{Pl}$.  This is different
from the usual seesaw mechanism, giving Dirac neutrinos instead of
Majorana neutrinos, yet their masses are naturally suppressed as
$m_{3/2} v / M_{Pl} \simeq 10$~meV.  This is actually closer to the
data of 9--50~meV than the standard GUT-based seesaw that predicts
$m_\nu \simeq 1$~meV.  Moreover, it is esthetically pleasing because
we do not need to rely on a new energy scale, {\it i.e.}\/ the
GUT-scale or the seesaw scale, to ``explain'' the neutrino mass.  It
is simply one of the Planck-scale suppressed operators.  

\subsection{Consistent Framework}

Recently, the successful electroweak symmetry breaking has been
demonstrated within the framework of the anomaly mediation with the
$D$-terms \cite{Kitano:2004zd}, which works particularly well with the
recently proposed Minimal Supersymmetric Fat Higgs
\cite{Harnik:2003rs} or a variant of the Next-to-Minimal
Supersymmetric Standard Model with extra vector-like quarks and
leptons \cite{Ibe}.  Therefore there are consistent models of anomaly
mediation with no apparent phenomenological problems, no flavor or CP
problems, no gravitino problem, and predicts naturally light Dirac
neutrinos.

\begin{figure}[htbp]
  \centering \includegraphics[width=\columnwidth]{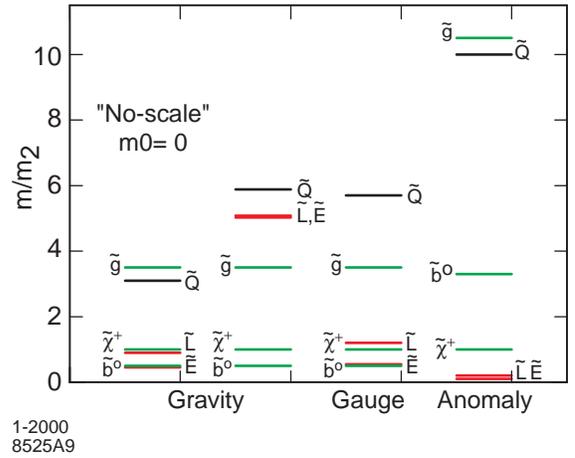}
  \caption{Sample SUSY spectra among various SUSY breaking mechanisms
  \cite{Peskin:2000ti}.} 
  \label{fig:SUSYspectra}
\end{figure}

\begin{figure}[htbp]
  \centering \includegraphics[width=\columnwidth]{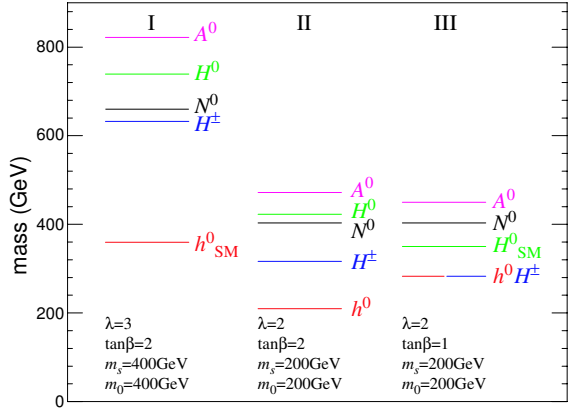}
  \caption{Sample Higgs spectra in a consistent model with
    anomaly-mediated supersymmetry breaking and the Fat Higgs in
    \cite{Kitano:2004zd}.}
  \label{fig:higgsspectrum7}
\end{figure}

The superparticle spectrum of the anomaly mediation is quite different
from many other frameworks (see Fig.~\ref{fig:SUSYspectra}), in
particular the peculiar ratio among the gaugino masses due to the
gauge beta functions.  The Higgs sector is also likely to be richer
than the standard MSSM with unusual mass spectra (see
Fig.~\ref{fig:higgsspectrum7}).  Future experiments, which require at
least the LHC and the ILC, but possibly also VLHC or CLIC, may well
verify such unusual superparticle spectra.

\subsection{Dirac Leptogenesis}

Having presented a model of naturally light Dirac neutrinos, an
obvious question arises: how do we understand the baryon asymmetry of
the universe?  The lepton number violation in Majorana neutrinos is
the crucial ingredient in the leptogenesis \cite{Fukugita:1986hr}.

Dirac leptogenesis overcomes this problem by the following simple
observation \cite{Dick:1999je}.  Recall that the Dirac neutrinos have
tiny Yukawa couplings, $m_\nu = Y_\nu v,$ $Y \simeq m_{3/2}/M_{Pl}
\simeq 10^{-13}$.  If this is the only interaction of the right-handed
neutrinos, thermalization is possible only by the process $N L
\rightarrow H W$ etc, and they do not thermalize for $T \gtrsim g^2
Y_\nu^2 M_{Pl} \sim 10$~eV.  (A similar but less dramatic point about
the electron Yukawa coupling was made in \cite{Campbell:1992jd}.)  At
this low temperature, obviously both $H$ and $W$ cannot be produced
and the thermalization is further delayed until $T_\nu \simeq m_\nu$
when neutrinos become non-relativistic.  Therefore the number of
left-handed and right-handed neutrinos are separately conserved
practically up to now.  We call them $L$ and $N$, respectively, and
the total lepton number is $L+N$.  The combination $L+N-B$ is strictly
conserved.

\begin{figure}[tbp]
  \centering \includegraphics[width=0.5\columnwidth]{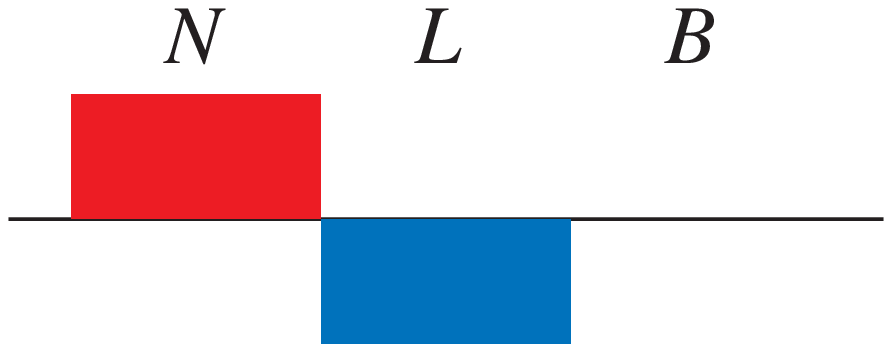}
  \includegraphics[width=0.5\columnwidth]{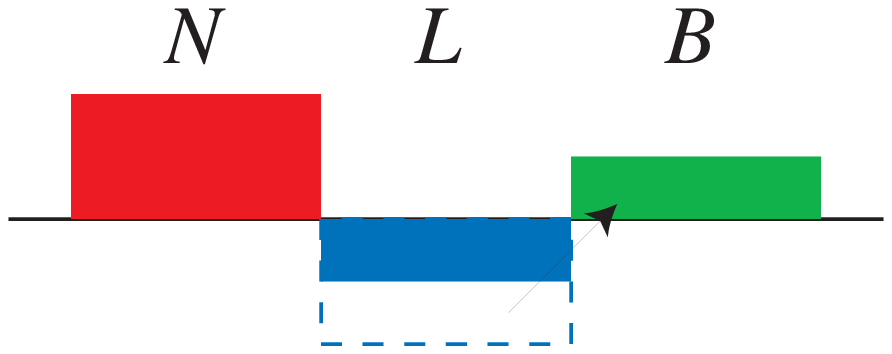}
  \includegraphics[width=0.5\columnwidth]{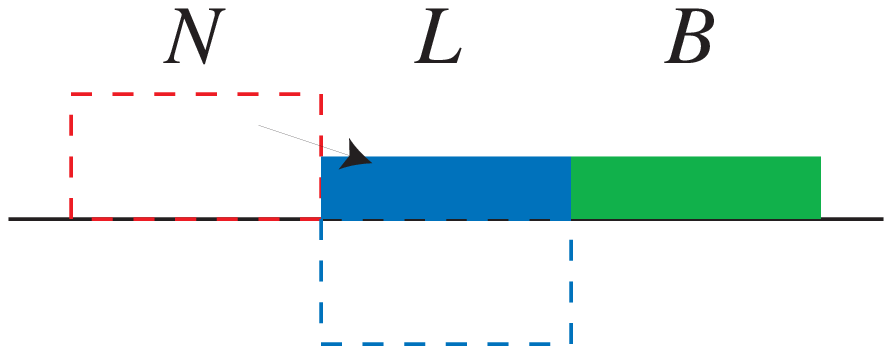}
  \caption{The evolution of the lepton asymmetry in Dirac leptogenesis
    models.  At the first stage, an asymmetry between the ordinary
    leptons and the right-handed neutrinos is created without
    lepton-number violation.  Then the asymmetry in the ordinary
    leptons is partially converted to the baryon asymmetry.  Finally,
    the right-handed neutrinos come in thermal equilibrium with the
    other leptons.  The net baryon and lepton asymmetries remain while
    the overall $B-L$ vanishes.}
  \label{fig:Dirac}
\end{figure}

Suppose the decay of a heavy particle produced an asymmetry $L_0 =
-N_0 \neq 0$.  The overall lepton number is conserved (see
Fig.~\ref{fig:Dirac}).  $N_0$ is frozen down to $T_\nu$.  On the other
hand, the lepton asymmetry $L_0$ is partially converted to the baryon
asymmetry via the standard model anomaly \cite{Kuzmin:1985mm}.
Following \cite{Harvey:1990qw}, the chemical equilibrium due to the
sphaleron leads to $B \simeq 0.35 (B-L_0) = 0.35 N_0 \neq 0$ and $L
\simeq -0.65 (B-L_0) = -0.65 N_0$.  After the electroweak phase
transition $T \lesssim 250$~GeV, the anomaly is no longer effective
and $B \neq 0$ is frozen.  Finally at $T_\nu$, $L$ and $N$ equilibrate
with the total lepton asymmetry $L+N_0 \simeq 0.35 N_0$.  In the end
there is a baryon asymmetry $B = (L+N) \simeq 0.35 N_0$.

The original paper \cite{Dick:1999je} introduced new electroweak
doublet scalar $\phi$ that has the same quantum numbers as the Higgs
doublets and Yukawa couplings $\phi L N$ and $\phi^* L E$.  Note that
the model is not supersymmetric and hence the scalar field $\phi$ have
Yukawa coupling with complex conjugation.  If there are two sets of
them, there is CP violation and their decays can create the asymmetry
$L = -N \neq 0$.  A supersymmetric generation \cite{Murayama:2002je}
would require two separate chiral superfields $\phi$ and $\bar{\phi}$,
and a superpotential coupling $\phi L N + \bar{\phi} L E + M_\phi
\phi\bar{\phi}$.  This works with the anomaly-mediated supersymmetry
breaking.

\subsection{(Nearly) Verifiable Framework}

\begin{figure}[htbp]
  \centering
  \includegraphics[width=\columnwidth]{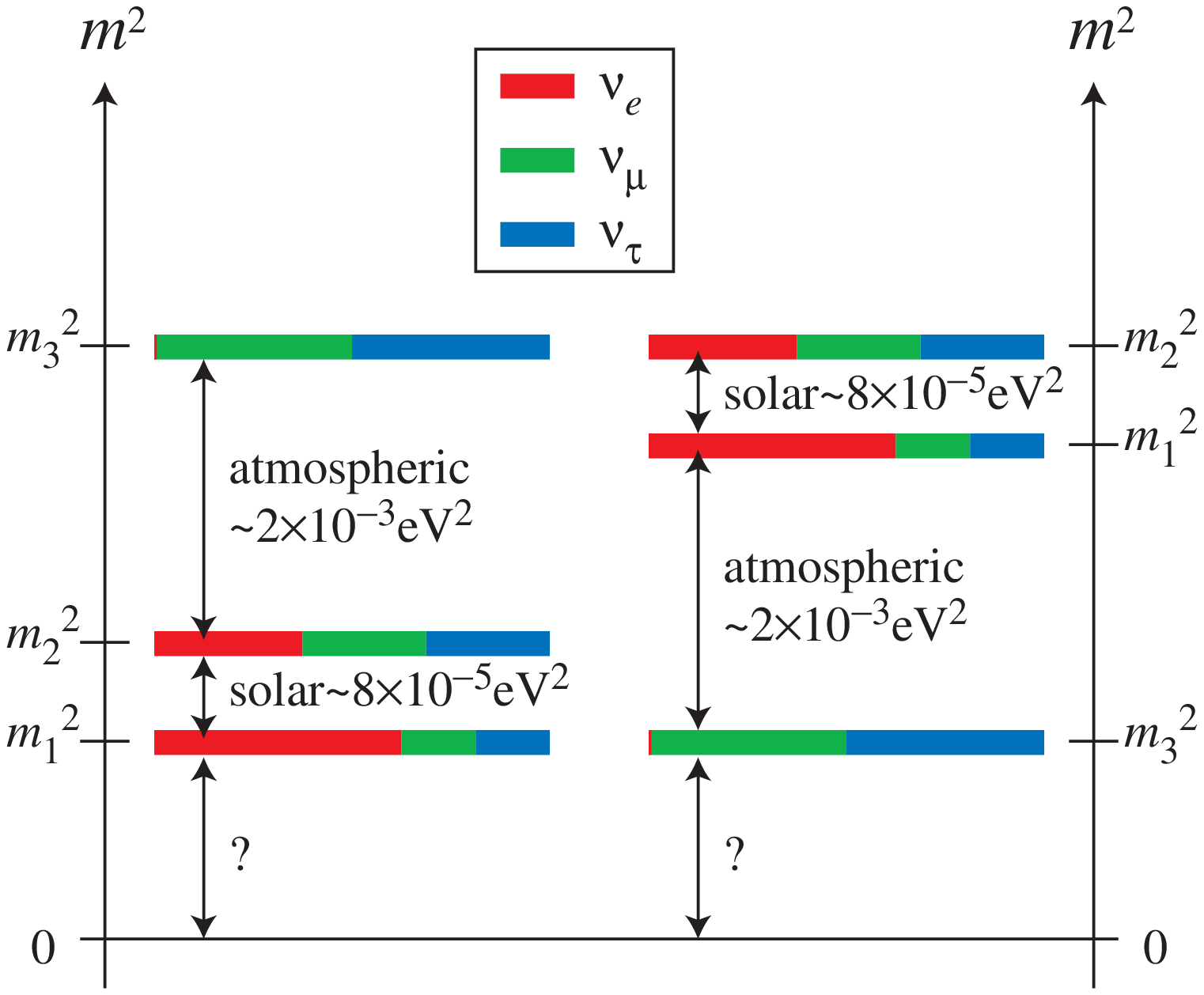}
  \caption{Two possible mass spectra of neutrinos.}
  \label{fig:spectrum}
  \centering
  \includegraphics[width=0.8\columnwidth]{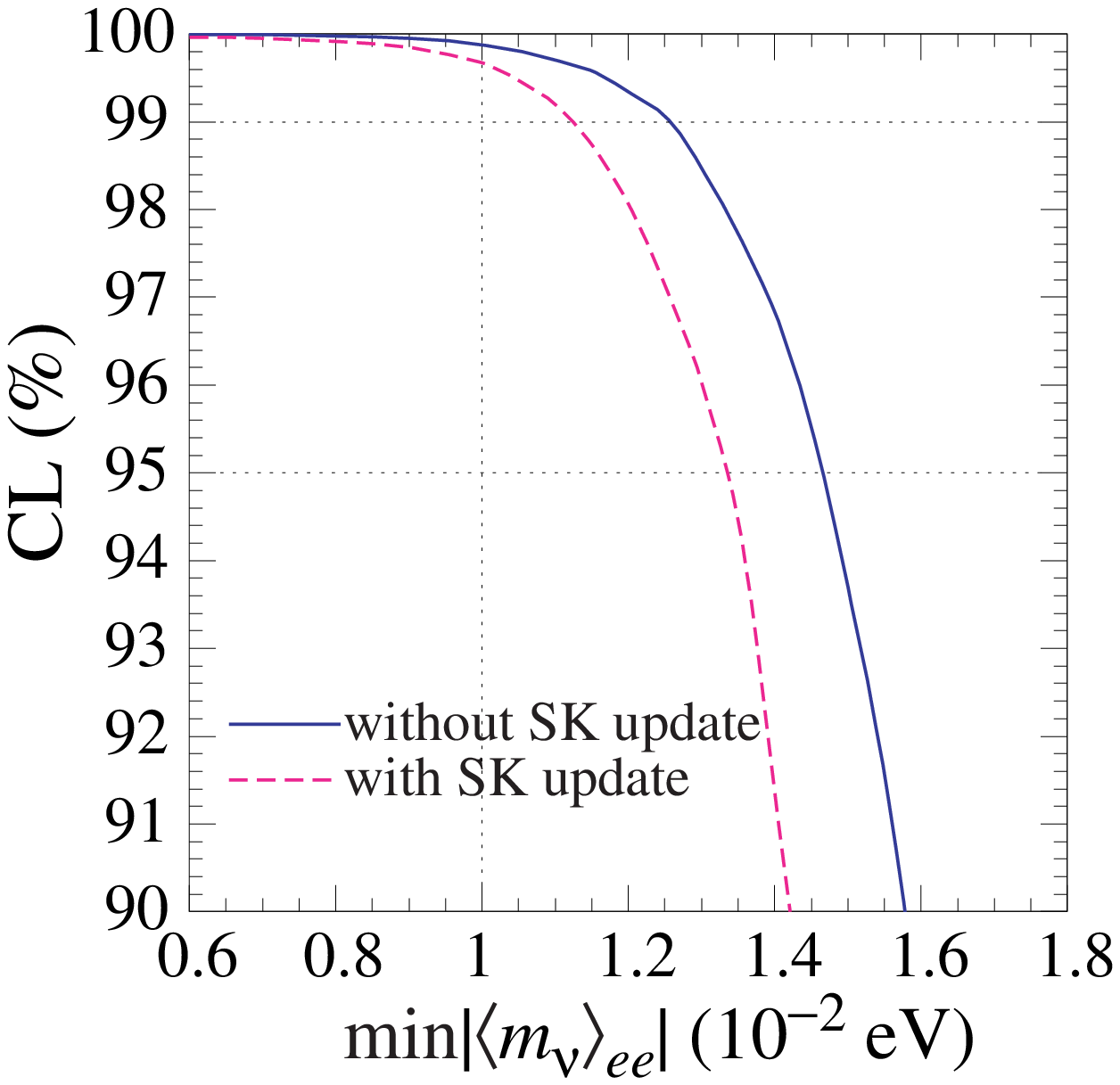}
  \caption{If the spectrum is inverted, there is a minimum size for
    the effective neutrino mass $|\langle m_{ee}\rangle|$ for the
    neutrinoless double-beta decay \cite{Murayama:2003ci}.}
  \label{fig:cl_new}
\end{figure}

Unlike the standard seesaw mechanism, the framework proposed here is
nearly verifiable.  (1) We need to find supersymmetry at the LHC, and
measure its spectrum at the ILC.  We can verify the mass spectrum of
anomaly mediation with $D$-terms.  (2) We may establish Dirac nature
of neutrinos.  For instance, if the long-baseline neutrino oscillation
experiments establish the inverted hierarchy of the (light) neutrino
spectrum (Fig.~\ref{fig:spectrum}), and if the neutrinoless double
beta decay sets a limit that $|\langle m_{ee}\rangle| = | \sum_{i=1}^3
U_{ei}^2 m_{\nu_i}| < 0.01$~eV, the Majorana neutrino hypothesis is
excluded (see Fig.~\ref{fig:cl_new}).\footnote{See, however,
  Ref.~\cite{Bahcall:2004ip} for precautions about the uncertainties
  in the nuclear matrix elements.}  Then the standard seesaw mechanism
is safely excluded, while there are strong experimental indications
for this framework.

\section{sMajorana}

Alternatives to the seesaw are not limited to Dirac neutrinos.  Here I
present a model where the light Majorana neutrinos are obtained yet
the right-handed neutrinos are present at the electroweak scale, and
hence offer the testability at collider experiments
\cite{Arkani-Hamed:2000bq,Borzumati:2000mc}.

Let us come back to the question ``why are neutrinos this light?''  In
the standard seesaw mechanism, the answer is that it is suppressed by
the GUT-scale, $m_\nu \sim v^2 / M_{GUT}$, where $v=176$~GeV is the
scale of the Higgs boson condensate.  Once one takes supersymmetry
seriously, this is not the only hierarchy we can use; the
supersymmetry breaking scale is supposed be similar (but slightly
higher) than the electroweak scale.  Indeed, it is possible to write
down a model where the neutrino mass is given by $m_\nu \sim
m_{SUSY}^2/M_{Pl} \sim 1$~meV for $m_{SUSY} \sim 1$~TeV.  In this
discussion, I do not assume the anomaly-mediated supersymmetry
breaking, but rather more conventional supergravity models.

In typical supergravity models, there is a field that breaks
supersymmetry at an intermediate scale $m_I \simeq (m_{SUSY}
M_{Pl})^{1/2}$.  It is quite plausible that the field also has a
supersymmetric expectation value of the comparable size, namely
$\langle X \rangle \simeq m_{I} + \theta^2 m_{I}^2$.  Then the
following Lagrangian
\begin{equation}
  \int d^2 \theta \frac{X}{M_{Pl}} L H_u N
  + \int d^4 \theta \frac{X^*}{M_{Pl}} N N
\end{equation}
picks up the expectation value of $X$ and gives
\begin{eqnarray}
  \lefteqn{
    \int d^2 \theta \sqrt{\frac{m_{SUSY}}{M_{Pl}}} L H_u N
  + m_{SUSY} \tilde{L} H_u \tilde{N}
  } \nonumber \\
  & &
  + \int d^2 \theta m_{SUSY} N N.
\end{eqnarray}
Therefore, the neutrino Yukawa coupling is given by $Y_\nu \simeq
\sqrt{m_{SUSY}/M_{Pl}} \simeq 10^{-7.5}$, while the right-handed
neutrino has the mass of order TeV.  There is a little seesaw when the
right-handed neutrinos are integrated out, giving the Majorana mass of
light neutrinos $m_\nu \simeq m_{SUSY}^2/M_{Pl} \simeq 1$~meV as
advertised.  The main difference from the standard seesaw is that the
right-handed neutrinos are light, allowing for direct experimental
tests.  The Lagrangian presented above can be natural with a $U(1)_R$
symmetry, under which charges are assigned as $N(2/3)$, $X(4/3)$,
$L(0)$, $H_u(0)$.

A very interesting term in the above Lagrangian is the trilinear
scalar coupling $m_{SUSY} \tilde{L} H_u \tilde{N}$.  Once the Higgs
boson acquires a VEV, it gives the left-right mixing mass-squared term
of $O(m_{SUSY} v)$, comparable to the left-left and right-right
sneutrino mass-squared terms.  Therefore, we expect the left-handed
sneutrino $\tilde{\nu}$ and right-handed sneutrinos $\tilde{N}$ to mix
with an $O(1)$ angle.  This is quite unique to this framework.  In
particular, when sneutrinos are produced at collider experiments, we
have a chance to see the mixture of the right-handed component,
proving that there are right-handed sneutrinos at the electroweak
scale.  This would be a clear evidence that the standard seesaw is not
at work.

The fact that the left-handed and right-handed sneutrinos mix
substantially is also very interesting for cosmology.  The sneutrino
can be a viable Dark Matter candidate.  For a normal left-handed
sneutrino, the annihilation cross section is too large to leave enough
abundance and/or the detection cross is too large and is already
excluded by the direct detection experiments (see, {\it e.g.}\/,
\cite{Falk:1994es}).  The mixture of left- and right-handed sneutrinos
evade both problems.  The annihilation cross section is suppressed by
$\sin^4 \theta$ where $\theta$ is the left-right mixing angle.

\begin{figure}[htbp]
  \centering
  \includegraphics[width=\columnwidth]{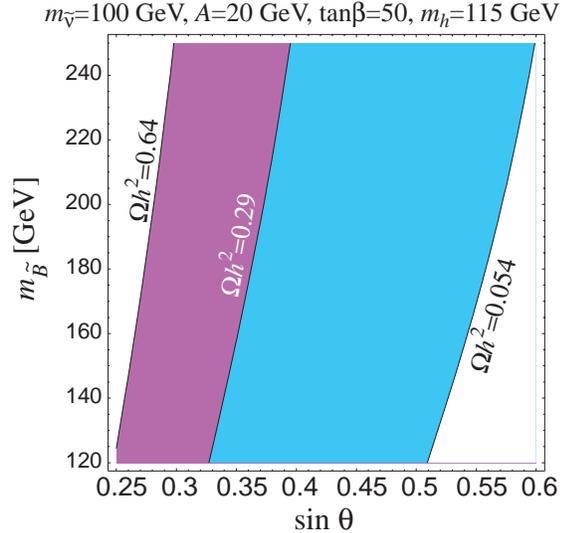}
  \caption{The cosmological abundance of the sneutrino dark matter as
    a function of the left-right mixing angle and the bino mass
    \cite{Arkani-Hamed:2000bq}.}
  \label{fig:Omega} 
\end{figure}

From the point of the direct detection experiments, there is another
highly suppressed operator that turns out to be interesting:
\begin{equation}
  \label{eq:2}
  \int d^4 \theta \frac{X X^* X^*}{M_{Pl}^3} N N
  \ni \frac{m_I^5}{M_{Pl}^3} \tilde{N} \tilde{N}.
\end{equation}
It violates the lepton number and mixes the right-handed sneutrino and
the anti-right-handed sneutrino.  The mixing between a scalar and its
anti-particle is reminiscent of the neutral kaon and $B$-meson
systems.  It gives the mass splitting between two mass eigenstates,
one CP-even and the other CP-odd, of the order of $\Delta m \simeq
(m_{SUSY}^3/M_{Pl})^{1/2} \simeq 100$~keV.  This number is particularly
interesting because it is roughly the kinetic energy of the dark
matter particles in our galactic halo $\frac{1}{2} m v^2 \simeq
\frac{1}{2} (100~{\rm GeV}) (10^{-3})^2 = 50$~keV.

It has been known that the lepton-number violation in the sneutrinos
has an important consequence on the direct detection experiments
because it will kill the diagonal coupling to the $Z$-boson.  The two
mass eigenstates, CP-even $\tilde{\nu}_+$ and CP-odd $\tilde{\nu}_-$,
can couple only {\it off-diagonally}\/,
$Z$-$\tilde{\nu}_+$-$\tilde{\nu}_-$ \cite{Hall:1997ah}.  This is a
simple consequence of the Bose symmetry: two identical bosons cannot
be in $P$-wave.  In direct detection experiments, the dominant
scattering process is the exchange of the virtual $Z$-boson between
the sneutrino and the nucleus.  However, it is not an elastic
scattering, but an {\it inelastic}\/ process that transforms the
sneutrino mass eigenstate $\tilde{\nu}_1$ to a heavier state
$\tilde{\nu}_2$.  Because the mass splitting is approximately the same
as the kinetic energy, the dark matter scattering cross section is
affected {\it kinematically}\/ by the lepton-number violation.  Only
a part of the phase space $v^2 \geq \Delta m \frac{m+m_A}{m m_A}$,
where $m_A$ is the mass of the nucleus, allows for the inelastic
scattering of the sneutrino kinematically.

\begin{figure}[htbp]
  \centering
  \includegraphics[width=0.5\columnwidth]{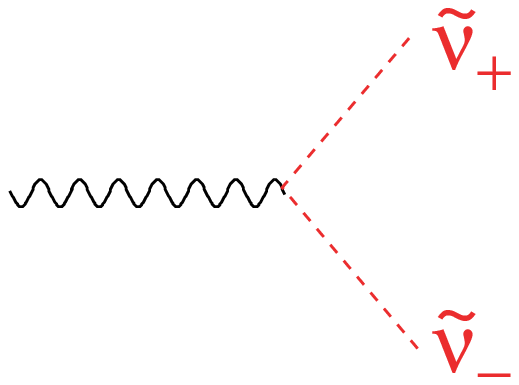}
  \caption{The coupling of the CP-odd and CP-even sneutrino mass
  eigenstates to the $Z$-boson is off-diagonal.} 
  \label{fig:Znu+nu-}
%% \end{figure}

%% \begin{figure}[htbp]
  \centering
  \includegraphics[width=\columnwidth]{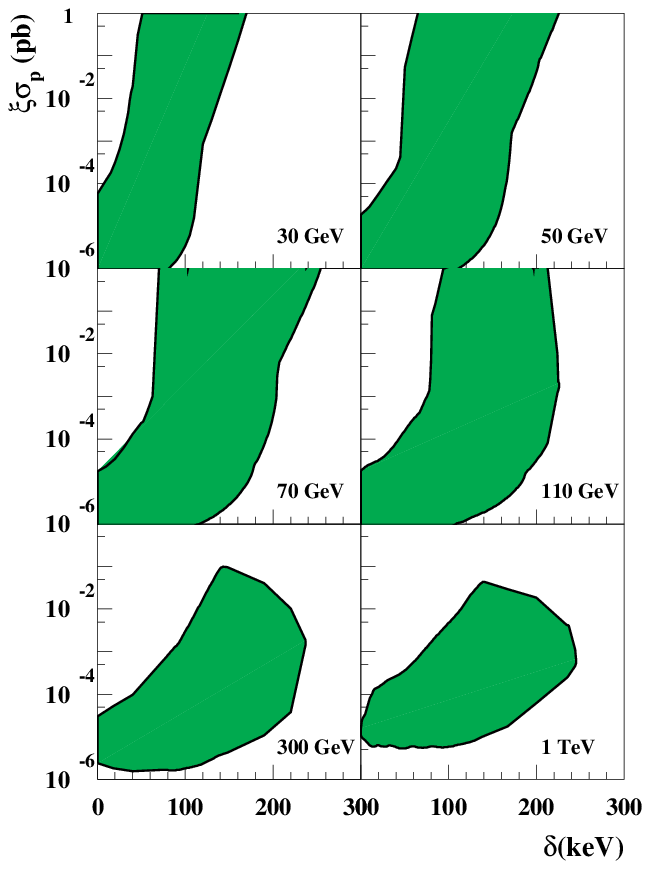}
  \caption{Preferred region of the parameter space for the inelastic
    Dark Matter by the DAMA experiment \cite{Bernabei:2003za}.  The
    horizontal axis $\delta$ is the mass splitting between the two
    mass eigenstates, while the vertical axis is the cross section on
    a nucleon assuming the spin-independent coherent matrix element.}
  \label{fig:DAMA}
\end{figure}

\begin{figure}[htbp]
  \centering
  \includegraphics[width=\columnwidth]{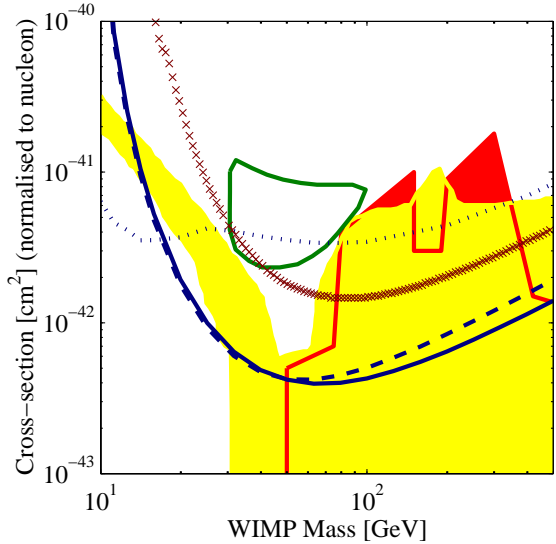}
  \caption{Limit on the elastic spin-independent coherent scattering
  cross section of Dark Matter from CDMS-II, together with the
  preferred region from DAMA \cite{Akerib:2004fq}.} 
  \label{fig:CDMS-II}
\end{figure}

This dark matter candidate can reconcile
\cite{Smith:2001hy,Tucker-Smith:2004jv} the claimed
6.3~$\sigma$evidence for the dark matter detection by the DAMA
experiment \cite{Bernabei:2003za} and the negative search by the
CDMS-II experiment \cite{Akerib:2004fq}.  The DAMA evidence is based
on the annual modulation of the event rate that is interpreted as a
consequence of the slight shifts in the phase space distribution of
Dark Matter particles due to the motion of the Earth relative to that
of the solar system inside the Milky Way galaxy.  The kinematic
selection of a part of the phase space due to the inelasticity
enhances the annual modulation effect.  Moreover, the heavier iodine
nucleus (NaI in DAMA) has the larger phase space available than the
lighter germanium (CDMS) and makes the signal larger in DAMA.  The
global fit to both data sets suggests a good fit and the mass
difference of about 100~keV as expected \cite{inelastic}.

The collider signature is also quite interesting.  Because the
sneutrino is the LSP, every superparticle decays eventually down to
the sneutrino state.  It can cause a very confusing situation in
interpreting the signal \cite{deGouvea:1998yp}.  The chargino decays
into a charged lepton and missing energy, normally associated with the
slepton signal.  The slepton decays into two jets and missing energy,
normally associated with the chargino signal.  The signature at the
LHC may be completely misinterpreted.  Sorting it out is not easy even
at the ILC, but it is possible in principle to choose the correct
interpretation by measuring the spins of superparticles for each
signal topology, their threshold behavior, their cross sections, decay
angle distributions, and azimuthal correlation between decay planes
\cite{LCWS2000}.  

\begin{figure}[htbp]
  \centering
  \includegraphics[width=0.6\columnwidth]{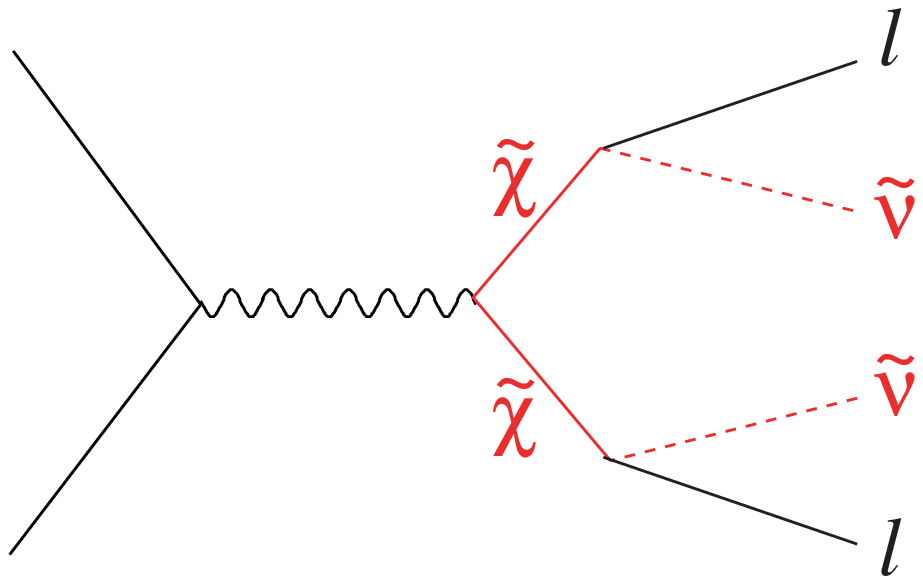}
  \includegraphics[width=0.6\columnwidth]{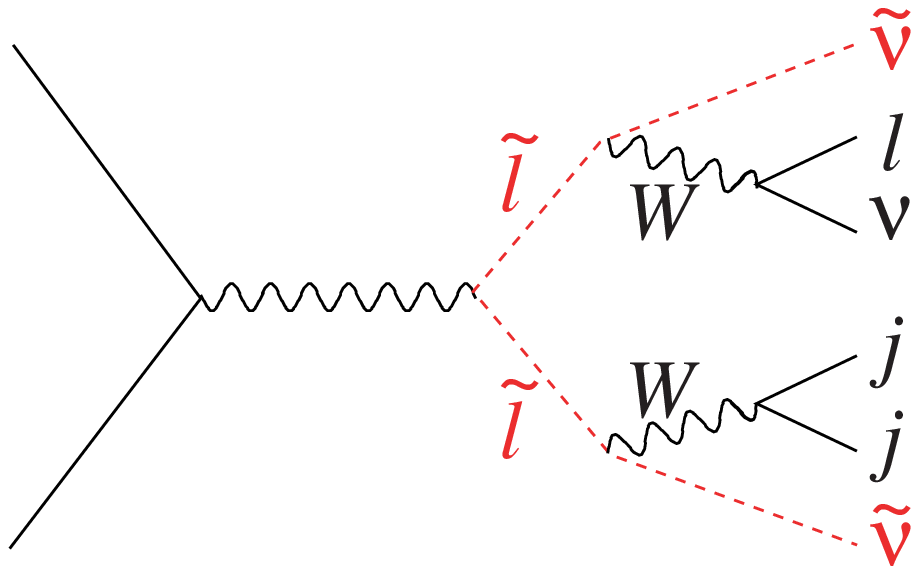}
  \caption{Unusual signatures of slepton and charginos that are
  interchanged from the conventional scenarios.} 
  \label{fig:confusing}
\end{figure}

\begin{figure}[htbp]
  \centering
  \includegraphics[angle=-90,width=0.9\columnwidth]{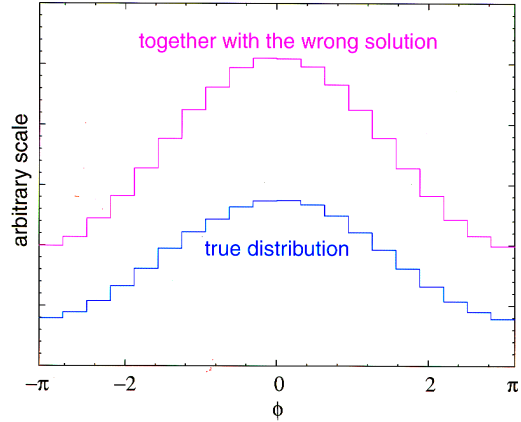}
  \caption{The azimuthal correlation between two decay planes from the
    chargino pair production in dilepton+missing topology.  Together
    with the clear two-body kinematics in the charged lepton energy
    distribution, it establishes that it is a spin-1/2 particle,
    excluding the slepton interpretation of the signal
    \cite{LCWS2000}.}
\end{figure}

Moreover, the measurement of the left-right mixing angle is possible
in the heavier sneutrino production \cite{ALCPG2004}.  It allows also
for the measurement of both sneutrino mass eigenvalues.  Together with
the charged slepton mass measurement, it should be possible to show
that the usual $D$-term formula does not work between the sneutrino
and the slepton, making the right-handed sneutrino mixing very clear.
Once the bino mass is measured, one can calculate
$\Omega_{\tilde{\nu}} h^2$ and compare it to the cosmological
measurement ({\it e.g.}\/, WMAP, Planck, weak lensing, etc).

\begin{figure}[htbp]
  \centering
  \includegraphics[width=0.8\columnwidth]{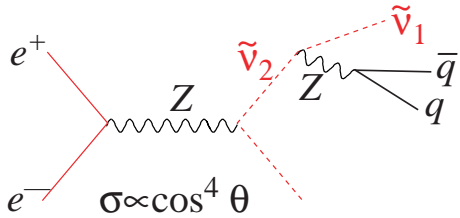}
  \caption{Measurement of the left-right mixing angle of sneutrinos at
  the Linear Collider \cite{ALCPG2004}.} 
  \label{fig:nu2}
\end{figure}

Once the right-handed sneutrinos below TeV are convincingly
established by the collider experiment, the standard seesaw mechanism
is unambiguously excluded.

This framework has been made even more attractive by recent works on
the explicit model of neutrino masses and mixings
\cite{March-Russell:2004uf}, the resonant leptogenesis
\cite{Hambye:2004jf,Chun:2004eq}, and the coincidence problem of
$\Omega_b$ and $\Omega_M$ \cite{Hooper:2004dc}.

\section{Conclusions}

Despite its attraction, the standard seesaw mechanism has many
problems.  The consistent anomaly mediation allows naturally light
Dirac neutrinos at the correct order of magnitude for the neutrino
mass, and solves the flavor, CP, and gravitino problems.  The
sMajorana model achieves a little seesaw with right-handed neutrinos
below TeV, giving rise to direct collider tests and inelastic Dark
Matter that reconciles DAMA and CDMS-II.  It is clear that it is
worthwhile pursing alternatives to the seesaw with a keen attention to
the testability.

Nonetheless the spirit of the seesaw lives: small neutrino mass is a
window to the physics beyond the Standard Model.

\begin{figure}[htbp]
  \centering
  \includegraphics[width=0.6\columnwidth]{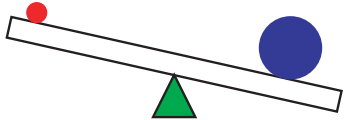}
  \caption{The spirit of the seesaw lives!}
  \label{fig:seesaw}
\end{figure}

\section*{Acknowledgments}

I thank the organizers of the seesaw25 workshop for the excellent
organization and exciting workshop, and in particular to Prof. Kenzo
Nakamura for his patience waiting for my manuscript.  This work was
supported by the Institute for Advanced Study, funds for Natural
Sciences, as well as in part by the DOE under the contract
DE-AC03-76SF00098 and in part by the NSF grant PHY-0098840.

\end{document}